\documentstyle[fleqn]{article} 
\begin{document}

\noindent
Title: ASSOCIATED LAM\'{E} EQUATION,PERIODIC POTENTIALS AND sl(2,R)

\noindent
 Author: ASISH GANGULY

\noindent
 Journal ref: published, Mod. Phys. Lett. A15(2000),1923

 \noindent
 Comment: 8 pages, no figure, tex file(version 2.09)

\begin{abstract}
We propose a new approach based on the algebraization of the Associated Lam\'{e} equation
\[-\psi''(x) + [ m(m+1)k^{2}sn^{2}x + \ell(\ell+1)k^{2}(cn^{2}x/dn^{2}x)]\psi(x) = E\psi(x)\]
within sl(2,R) to derive the corresponding periodic potentials. The band edge eigenfunctions and energy 
spectra are explicitely obtained for integers m,$\ell$. We also obtain the explicit expressions of the 
solutions for half-integer m and integer or half-integer $\ell$.
\end{abstract}
 \pagebreak

Periodic potentials have attracted considerable attention of late[1-8]. In the study of
quantized Neumann problem,generalized Lam\'{e} equation was obtained[1] based on the approach
of variables separation. A model on periodic potentials was constructed[2a] in the framework of
supersymmetric quantum mechanics. A relation with Lam\'{e} equation was established in Ref~2b
and it was shown that some families of Lam\'{e} potentials turn out to be self-isospectral. A geometric 
approach on elliptic finite-gap potentials was also studied in Ref~3. Moreover,Andrianov~et~al~[~4~] also 
obtained several elliptic potentials for some integrable two-dimentional systems in connection with Lax 
method.

Several periodic potentials such as Razavy potential[5],Lam\'{e} potential or the  associated 
Lam\'{e} potential[6] are known to be quasi-exactly solvable(QES). A study of Lie-algebraic approach has 
been made extensively to find the hidden symmetry algebra of the QES potentials. It is now known that for a 
one-dimentional QES Hamiltonian,hidden symmetry is sl(2,R). Attempts have been made to find a Lie-algebraic
representation of several QES periodic potentials[7,8]. The class of QES periodic Hamiltotians representable
in terms of sl(2,R) generators is known as algebraic QES. The QES property of an algebraic potential
immediately follows from the fact that the matrix representing the action of Hamiltonian has finite block 
structure so that by diagonalizing the matrix, a part of the sprectrum can be computed in a pure algebraic 
way. Unfortunately, the list of such algebraic QES periodic potentials is very short. In this letter we 
propose an algebraization of the associated Lam\'{e} equation based on the underlying group SL(2,R). To the 
best of our knowledge such a study has not been undertaken so far in the literature. Interestingly our 
solutions include as a subset those discussed in Ref~8.

To begin with, let us consider the following differential realization of the sl(2,R) 
generators $T^{\pm},T^{0}$
\begin{equation}
T^{+} = \xi^{2}\partial_{\xi} - n\xi, \qquad  T^{0}=\xi\partial_{\xi} - \frac{1}{2}n, \qquad 
                                                                                T^{-}= \partial_{\xi},
\end{equation}
where n is a non-negative integer. These generators act on polynomials in real variable $\xi$ of deg
$\leq $ n. The sl(2,R) algebra is given by the commutation relations
\begin{equation}
     [T^{+},T^{-}]  = -2T^{0} , \qquad  [T^{0},T^{\pm}] = \pm T^{\pm}
\end{equation}
Now the general quadratic combination of the generators induce the gauged Hamiltonian
\begin{equation}
H_{G} = - \sum_{a,b=0,\pm} C_{ab}T^{a}T^{b} - \sum_{a=0,\pm} C_{a}T^{a} - d,
\end{equation}
Using (1), $H_{G}$ can be represented as
\begin{equation}
H_{G}(\xi) = -\sum_{i=2}^{4} B_{i}(\xi)\partial_{\xi}^{i-2} ,
\end{equation}
where $B_{i}(\xi)$ are the ith degree polynomial in variable $\xi$ given by
\begin{eqnarray}
B_{4}(\xi) & = & C_{++}\xi^{4} + 2C_{+0}\xi^{3} + C_{00}\xi^{2} + 2C_{0-}\xi + C_{--},\nonumber \\
B_{3}(\xi) & = & 2(1-n)C_{++}\xi^{3} +\{ 3(1-n)C_{+0}+C_{+}\}\xi^{2} + \{ (1-n)C_{00}+C_{0}\}\xi \nonumber \\
           &   & \mbox{} +(1-n)C_{0-} + C_{-} , \nonumber \\
B_{2}(\xi) & = & n(n-1)C_{++}\xi^{2} + n\{ (n-1)C_{+0} - C_{+}\}\xi 
                                                           + \frac{n^{2}}{4}C_{00} - \frac{1}{2}nC_{0} + d .
\end{eqnarray}
Note that the numerical parameters $\{C_{i,j}\}$ are symmetric with $C_{+-} = C_{-+} = 0$ and d is a 
suitably chosen constant. A coordinate transformation of the type
\begin{equation}
x(\xi) = \int^{\xi} \frac{d\tau}{\sqrt{B_{4}(\tau)}}
\end{equation}
converts $H_{G}$ to the form
\begin{equation}
H_{G}(x) = -\partial_{x}^{2} +\left [ \frac{B_{4}^{'}-2B_{3}}{2 \surd B_{4}}\right ]_{\xi=\xi(x)} \partial_{x} 
                                                   - \left [ B_{2} \right ]_{\xi=\xi(x)} , 
\end{equation}
where the prime denotes derivative with respect to $\xi.$

On the other hand, it is well known that the Schr\"{o}dinger equation
\begin{equation}
H(x)\psi(x) \equiv [-\partial_{x}^{2} + V(x)]\psi(x) = E\psi(x)
\end{equation}
can be gauge-transformed using an imaginary phase transformation
\begin{equation}
\psi(x) = e^{-\int {\cal A}(x)dx}\chi(x)
\end{equation}
to the form
\begin{equation}
H_{G}(x)\chi(x) \equiv [-\partial_{x}^{2} +2 {\cal A} \partial_{x} + \frac{d{\cal A}}{dx} 
                                                                           -{\cal A}^{2} + V]\chi(x),
\end{equation}
where $ {\cal A}(x)$ is some gauge function.

Comparison between Eqs. \ (7) and (10) reveal
\begin{equation}
{\cal A}(x) = \left [\frac{B_{4}^{'}-2B_{3}}{4 \sqrt{B_{4}}}\right ]_{\xi=\xi(x)} , \;
V(x) =\left [{\cal A}^{2} - \frac{d{\cal A}}{dx} - B_{2} \right ]_{\xi=\xi(x)} .
\end{equation}
The Schr\"{o}dinger potential V(x) can be written in terms of $B_{i}(\xi)$ as follows :
\begin{equation}
V(x) = \left [ \frac{(B_{4}^{'} - 2B_{3})(3B_{4}^{'} -2B_{3})}{16B_{4}} - \frac{1}{4}(B_{4}^{''}- 2B_{3}^{'} +
               4B_{2})\right ]_{\xi=\xi(x)} .
\end{equation}

We wish to remark that the family of potentials (12) is reducible [9] to a class of elliptic potentials for
a general choice of $B_{4}(\xi)$ and the construction of Hamiltonian~(3) ensures that these potentials 
 form a quasi-exactly solvable system.  Indeed, for the choice
\begin{equation}
B_{4}(\xi) = (1 +\xi^{2})(1 + k^{'^{2}}\xi^{2}) 
\end{equation}
Eq. \ (6) can be inverted as
\begin{equation}
                           \;    \xi = sn \, x/cn \, x ,
\end{equation}
where $sn \, x \equiv sn(x,k),cn \, x \equiv cn(x,k),dn \, x \equiv dn(x,k)$ are three Jacobian elliptic
functions of real modulus $k(0< k^{2} < 1)$ and $k^{'^{2}} = 1- k^{2}$ is the complementary
modulus.

Given $B_{4}(\xi)$ as above, the form of (12) becomes
\begin{equation}
V(x) = P \, sn^{2}x + Q \, sn \, xcn \, x + R \frac{sn \, xcn \, x}{dn^{2}x} + S \frac{cn^{2}x}{dn^{2}x} ,
\end{equation}
where
\begin{displaymath}
P = \frac{k^{2}}{4}n(n+2) - \frac{C_{0}}{2}(n+1)+ \frac{1}{4k^{2}}[C_{0}^{2} \, - (C_{+} - C_{-})^{2}],
\end{displaymath}
\begin{displaymath} 
Q = \frac{1}{2k^{2}}(C_{+} - C_{-})[ k^{2}(n+1) - C_{0}] , 
\end{displaymath}
\begin{displaymath}
R = \frac{1}{2k^{2}}(C_{+} - k^{'^{2}}C_{-})[ k^{2}(n+1) + C_{0}] , 
\end{displaymath}
\begin{equation}
S = \frac{k^{2}}{4}n(n+2) + \frac{C_{0}}{2}(n+1) + \frac{1}{4k^{2}}[C_{0}^{2} - \frac{1}{k^{'^{2}}}
                                                              (C_{+}- k^{'^{2}}C_{-})^{2}] ,
\end{equation}
and d is chosen as
\begin{equation}
d = \frac{1}{4k^{2}}[C_{-}^{2} - (C_{0}^{2} + 2C_{+}C_{-}) + (\frac{C_{+}}{k^{'}})^{2}] - \frac{n(n+2)}{2} .
\end{equation}

We now turn to the associated Lam\'{e} equation
\begin{equation}
  -\psi^{''}(x) + [m(m+1)k^{2}sn^{2}x +\ell(\ell+1)k^{2}\frac{cn^{2}x}{dn^{2}x}]\psi(x) = E\psi(x) ,
\end{equation}
where $\ell,m\in R$ and $m\geq\ell$ without loss of generality. Equation(18) reduces to the ordinary Lam\'{e}
equation when either of $\ell$ and $m$ takes a value 0 or -1. Comparing the Schr\"{o}dinger equation~(8)
 having the potential(15) with (18) shows the following correspondence:
\begin{equation}
  P = k^{2}m(m+1) \, , \; Q = R = 0 \, , \; S = k^{2}\ell(\ell+1) .
\end{equation}

Four nontrivial solutions emerge which are summarized bellow
\begin{eqnarray}
n=m+\ell:         & C_{+} = C_{-} = 0,               & C_{0} = k^{2}(\ell-m) ,\\
n=m-\ell-1:       & C_{+} = C_{-} = 0,               & C_{0} = -k^{2}(\ell+m+1) ,\\
n=m-\frac{1}{2} : & C_{+} = C_{-} =ik^{'}(2\ell+1),  & C_{0} = -k^{2}(m+\frac{1}{2}),\\ 
n=m-\frac{1}{2} : & C_{+}= C_{-} =-ik^{'}(2\ell+1),  & C_{0}= -k^{2}(m+\frac{1}{2}).
\end{eqnarray}

The Schr\"{o}dinger Hamiltonian $H_{G}(x)$ can be expressed from(8) and (9) as
\begin{equation}
                H(x) =\left[ \mu(\xi)H_{G}\frac{1}{\mu(\xi)}\right]_{\xi=\xi(x)} , 
\end{equation}
where the gauge factor \( \mu((\xi(x)) = exp \, [-\int{\cal A}(x) \,dx] \) and the gauge Hamiltonian 
 $H_{G}$ can be easily computed from Eqs (3) and (11) corresponding to each of the 
 four algebraizations(20)-(23).

 The band edge wave functions and energy  spectra of the associated Lam\'{e} potential may now be explicitely
 formulated using the  techniques of Bender and Dunne[10].  We do not give the details of our calcutations
 which will be  communicated elsewhere[11].  Here we briefly outline our results.  In the following two
  particular  cases are considerd. 

 \noindent
 {\bf  Case 1. \ $\mathbf{m}$ and $\mathbf{\ell}$ are both non-negative integer}

 \noindent
 Here $m$ takes values 0,1,2, \ldots and for each value of $m,\ell$ is restricted to take ($m+1$)
 values $0,1, \ldots ,m$. The lowest value of $\ell$ gives the ordinary Lam\'{e} potential. This case 
 corresponds to the algebraizations(20) and(21). First algebraization gives $m+\ell+1$ eigenstates while the 
 second algebraization gives $m-\ell$ eigenstates. Hence when $m$   and $\ell$ are both 
 non-negative integers ($m\geq\ell$), there are $m$ bound bands followed by a continuum band and associated 
 Lam\'{e}  Hamiltonian possesses ($2m+1$) band edge eigenstates at the lower and upper edges of each band. 
 The latter  generates two distinct families of orthogonal polynomials for the energy variable E. In the 
 following  examples $\phi_{r}(x)$ and $e_{r}$ denote ordered levels of eigenstates and energy spectra.

 \noindent
 {\bf Examples.}

 \noindent
(a) $\mathbf m=1$
 \begin{equation}
 (i) \ell = 0:\mbox{Lam\'{e} potential} V(x) = 2k^{2}sn^{2}x
 \end{equation}  
 \[ \begin{array}{lll}
                      \phi_{0}(x) = dnx ,& \phi_{1}(x) = cnx ,& \phi_{2}(x) = snx 
 \end{array} \]
 \[ \begin{array}{lll}
                      e_{0} = k^{2} ,   & e_{1} = 1,          & e_{2} = 1+ k^{2} .
 \end{array} \]
 \begin{equation}
 (ii) \ell = 1:\mbox{Associated Lam\'{e} potential} V(x) = 2k^{2}sn^{2}x + 2k^{2}(cn^{2}x/dn^{2}x)
 \end{equation}
 \[ \begin{array}{ll}
                  \phi_{0,1}(x) = dnx \pm (k^{'}/dnx),& e_{0,1} = 2+k^{2}\mp 2k^{'} ,
 \end{array} \]
 \[ \begin{array}{ll}
                  \phi_{2}(x) = (sn cnx)/dnx ,    & e_{2} = 4
 \end{array} \] \vspace{.3cm}

\noindent
(b)$\mathbf m = 2$
 \begin{equation}
 (i)\ell =0:\mbox{Lam\'{e} potential} V(x) = 6k^{2}sn^{2}x
 \end{equation}
\begin{displaymath}
 \phi_{0,4}(x)= 3dn^{2}x + k^{2}-2\pm \sqrt{k^{4}-k^{2}+1},\quad e_{0,4} = 2(1+k^{2}\mp \sqrt{k^{4}-k^{2}+1} ,
 \end{displaymath}
 \[ \begin{array}{lcl}
             \phi_{1}(x) = cnx dnx , & \phi_{2}(x) = snx dnx , & \phi_{3}(x) = snx cnx , 
 \end{array} \]
 \[ \begin{array}{lcl}
             e_{1} = 1+ k^{2} ,             & e_{2} = 1+ 4k^{2} ,            & e_{3} = 4+k^{2}.
 \end{array} \]
 \begin{equation}
 (ii) \ell = 1:\mbox{Associated Lam\'{e} potential} V(x) = 6k^{2}sn^{2}x + 2k^{2}(cn^{2}x/dn^{2}x)
 \end{equation}
 \[ \begin{array}{ll}
 \phi_{0}(x) = dn^{2}x , & e_{0} = 4k^{2} , 
 \end{array} \]
 \[ \begin{array}{ll}
 \phi_{1,3}(x) = cnx[3dn^{2}x-1\pm \sqrt{4-3k^{2}}]/dnx,& e_{1,3} = 5+k^{2}\mp 2\sqrt{4-3k^{2}}, 
 \end{array} \]
 \[ \begin{array}{ll}
 \phi_{2,4}(x)=snx[3dn^{2}x-k^{'^{2}}\pm \sqrt{k^{4}-5k^{2}+4}]/dnx ,&
                                               e_{2,4}=5+2k^{2}\mp 2\sqrt{k^{4}-5k^{2}+4}.
 \end{array} \]
 \begin{equation}
 (iii) \ell = 2:\mbox{Associated Lam\'{e} potential} V(x) = 6k^{2}sn^{2}x + 6k^{2}(cn^{2}x/dn^{2}x)
 \end{equation}
 \[ \begin{array}{ll}
 \phi_{0,4}(x) = [1-\eta_{\mp}(k)sn^{2}x +\{\eta_{\mp}(k) - k^{2}\}sn^{4}x]/dn^{2}x, &
                                                        e_{0,4} = 2\eta_{\mp}(k) + 4k^{2} , 
 \end{array} \]
 \[ \begin{array}{ll}
 \phi_{1}(x) = [1- 2sn^{2}x + k^{2}sn^{4}x]/dn^{2}x , & e_{1} = 4(1+k^{2}) , 
 \end{array} \]
 \[ \begin{array}{ll}
 \phi_{2,3}(x) = snx cnx [1+(\pm k^{'}-1)sn^{2}x]/dn^{2}x , & e_{2,3} = 10 + k^{2}\mp 6k^{'} , 
 \end{array} \]
 \begin{displaymath}
 \mbox{where} \qquad \eta_{\pm}(k) = 4 - k^{2}\pm \sqrt{k^{4} - 16k^{2} + 16}.
 \end{displaymath}

 Similarly, higher integer values of $m$ can be dealt with. Let us point out that the Lam\'{e} potential
 corresponding to $\ell = 0$ in the above examples was obtained algebraically in Ref ~8. However, to the 
 best of our knowledge, the algebraic approach for the remaining solutions are new. \vspace{1.3cm}

 \noindent
{\bf Case 2. $\mathbf{m}$ is half an odd positive integer and $\mathbf{\ell}$ is either a non-negative \\ 
               integer or half an odd positive integer }

\noindent
Here $m$ takes values $(1/2),(3/2), \ldots$ and for each values of $m,\ell$ is restricted to take 
 $(2m+1)$  values $0,(1/2),1,(3/2), \ldots ,m$. The lowest value of $\ell$ gives the ordinary Lam\'{e}
 potential. Each of the two algebraizations(22) and (23) gives $[m+(1/2)]$ eigenstates, conjugate to each 
other, generating the same family of orthogonal polynomials for the energy variable E. This implies that 
energies are doubly degenerate and $2m+1$ eigenstates are given by real and imaginary parts. Thus there are 
$[m+(1/2)]$ chracteristic values of E for each of which we obtain two linearly independent solutions. In the 
following examples the parenthesized superscript in the eigenstates indicates the degeneracy of the 
eigenvalue.

\noindent
{\bf Example.}    

 \noindent
 (a) $\mathbf{m=}\frac{\mathbf 1}{\mathbf 2}$
 \begin{equation}
 (i) \ell = 0:\mbox{Lam\'{e} potential} V(x) = \frac{3}{4}k^{2}sn^{2}x
 \end{equation}
 \( \begin{array}{lll}
 \phi_{0}^{(1)}(x)=\sqrt{dn \, x+cn \, x}, & \phi_{0}^{(2)}=sgn(sn \, x)\sqrt{dn \, x-cn \, x},  &
                                                                                    e_{0}=(1+k^{2})/4 .
 \end{array} \)
 \begin{equation}
 (ii) \ell=\frac{1}{2}:\mbox{Associated Lam\'{e} potential} V(x)=\frac{3}{4}k^{2}sn^{2}x
                                                                   +\frac{3}{4}k^{2}(cn^{2}x/dn^{2}x)
 \end{equation}
 \( \begin{array}{lcl}
 \phi_{0}^{(1)}(x)=cn \, x/ \sqrt{dn \, x},& \phi_{0}^{(2)}(x)=sn \, x/ \sqrt{dn \, x},& e_{0}=1+(k^{2}/4).
 \end{array} \) \vspace{.3cm}

 \noindent
 (b) $\mathbf{m=}\frac{\mathbf 3}{\mathbf 2}$
 \begin{equation}
 (i) \ell=0:\mbox{ Lam\'{e} Potential} V(x)=\frac{15}{4}k^{2}sn^{2}x
 \end{equation}
 \[ \begin{array}{l}
 e_{0,1}=5(1+k^{2})/4 \, \mp \sqrt{k^{4}-k^{2}+1}, 
 \end{array} \]
 \[ \begin{array}{l}
 \phi_{0,1}^{(1)}(x)=\sqrt{(dn \, x+cn \, x)[k^{2}cn \, x+\alpha_{\pm}(k)dn \, x]}, 
 \end{array} \]
 \[ \begin{array}{l}
 \phi_{0,1}^{(2)}(x)=sgn(sn \, x)\sqrt{(dn \, x-cn \, x)[k^{2}cn \, x-\alpha_{\pm}(k)dn \, x]},
 \end{array} \]
 \begin{displaymath}
 \mbox{where}\qquad \alpha_{\pm}(k)=k^{'^{2}}\pm \sqrt{k^{4}-k^{2}+1}.
 \end{displaymath}

The above $\ell=0$ cases namely, those for the potentials(30) and (32), have already been reported in Ref 8.
\begin{equation}
 (ii)\ell=1/2:\mbox{Associated Lam\'{e} potential} V(x)=\frac{15}{4}k^{2}sn^{2}x+
                                                                  \frac{3}{4}k^{2}(cn^{2}x/dn^{2}x)
 \end{equation}
 \[ \begin{array}{lll}
 \phi_{0}(x)=dn^{3/2}x, & e_{0}=9k^{2}/4, &           
 \end{array} \]
 \[ \begin{array}{lll}
 \phi_{1}^{(1)}(x)=[2sn^{2}x-1]/ \sqrt{dn \, x}, & \phi_{1}^{(2)}(x)=[sn \, xcn \, x]/ \sqrt{dn \, x},
                                                                                        & e_{1}=4+(k^{2}/4). 
 \end{array} \]
 \begin{equation}
 (iii) \ell=1:\mbox{Associated Lam\'{e} potential} V(x)=\frac{15}{4}k^{2}sn^{2}x+2k^{2}(cn^{2}x/dn^{2}x)
  \end{equation}
 \[ \begin{array}{c}
 e_{0,1}=\frac{3}{2}k^{2}+\frac{13}{4} \mp \sqrt{k^{4}+9k^{'^{2}}}, 
 \end{array} \]
 \[ \begin{array}{l}
 \phi_{0,1}^{(1)}(x)=\sqrt{(dnx+cnx)[\beta_{\mp}(k)dn^{2}x+\gamma_{\mp}(k)cnxdnx+\delta_{\mp}(k)]/dnx}, 
 \end{array} \]
 \[ \begin{array}{l}
 \phi_{0,1}^{(2)}(x)=\sqrt{(dnx-cnx)[\beta_{\mp}(k)dn^{2}x-\gamma_{\mp}(k)cnxdnx+\delta_{\mp}(k)]/dnx} ,
 \end{array} \]
 \begin{displaymath}
 \mbox{where}\qquad \beta_{\pm}(k)=-7k^{4}+72 K^{2} - 96 \pm 8(k^{2}-4)\sqrt{k^{4}+ 9k^{'^{2}}},
 \end{displaymath}
 \begin{displaymath}
 \qquad \qquad \gamma_{\pm}(k)= 8k^{2}[- 3k^{2}+ 6\pm 2\sqrt{k^{4}+ 9k^{'^{2}}},
 \end{displaymath}
 \begin{displaymath}
 \qquad \qquad \delta_{\pm}(k) = 48k^{4} - 144k^{2} + 96\pm 32k^{'^{2}}\sqrt{k^{4} + 9k^{'^{2}}} .
 \end{displaymath}
 Thus from our algebraic approach we can find a new QES periodic potential(34) whose solutions can be written 
 analytically.
 \begin{equation}\
 (iv)\ell = 3/2:\mbox{Associated Lam\'{e} potential} V(x)=\frac{15}{4}k^{2}sn^{2}x
                                          +\frac{15}{4}k^{2}(cn^{2}x/dn^{2}x)
 \end{equation}
 \[ \begin{array}{l}
                    e_{0,1}= 5+\frac{5}{4}k^{2} \mp \sqrt{k^{4}-16k^{2}+16},
 \end{array} \]
 \[ \begin{array}{l}
                  \phi_{0,1}^{(1)}(x)=cnx[\phi_{\pm}(k)sn^{2}x+ 2k^{2}+\epsilon_{\pm}(k)]/(dn \, x)^{3/2} ,
 \end{array} \]
 \[ \begin{array}{l}
                  \phi_{0,1}^{(2)}(x)=snx[\rho_{\pm}(k)sn^{2}x +3\epsilon_{\pm}(k)]/(dn \, x)^{3/2} ,
 \end{array} \]
 \begin{displaymath}
 \mbox{where}\qquad \phi_{\pm}(k)= 3k^{4}-20k^{2}+16\pm(3k^{2}-4)\sqrt{k^{4}-16k^{2}+16} ,
 \end{displaymath}
 \begin{displaymath}
 \qquad \qquad \rho_{\pm}(k)=2k^{4}-12k^{2}+16\pm (k^{2}-4)\sqrt{k^{4}-16k^{2}+16} ,
 \end{displaymath}
 \begin{displaymath}
 \qquad \qquad \epsilon_{\pm}(k)=3k^{2}-4\pm \sqrt{k^{4}-16k^{2}+16} .
 \end{displaymath}

 Proceed in this way we can determine the potentials corresponding to $m=(5/2),(7/2), \ldots$ We may mention
 that the results of the cases $\ell=1/2,1,3/2, \ldots ,m$ given above using our algebraic approach are new.

To conclude, we have presented in this work new algebraizations of the associated Lam\'{e} equation. Within
 sl(2,R) these have enabled us to determine a general class of periodic potentials which include the ones 
 derived recently [8]. \vspace{.5cm}

\noindent
 {\bf Acknowledgment} \vspace{.2cm}

\noindent
I would like to thank Dr B Bagchi for guidance.

 \pagebreak

 \noindent
 {\bf References} 

\noindent
 \begin{tabbing}
 1. \= (a) \= tab setting                                      \kill
 1. \>     \> D Gurarie, J Math Phys {\bf 36},5355(1995).  \\ 
 2. \> (a) \> G Dunne and J Mannix, Phys Lett {\bf B428}, 115(1998),  \\
    \> (b) \> G Dunne and J Feinberg, Phys Rev {\bf D57}, 1271(1998).  \\
 3. \> (a) \> A R Its and V B Matveev, Theor Mat Fiz {\bf 23} , 51(1975) (in Russian) , \\
    \> (b) \> A Treibich and J L Verdier, Compt Rend Acad Sci Paris  {\bf 311},31(1990),  \\
    \> (c) \> see also the book: E D Belokolos, A I Bobenko, V Z Enol'skii and A R Its, \\
    \>     \> Algebro-geometric Approach to Nonlinear Integrable  equations \\
    \>     \> (Springer-Verlag,1994), Sec 7.7.                                 \\
 4. \>     \> A A Andrianov, M V Iof\mbox{}fe and D N Nishnianidze,J Phys {\bf A32} ,4641(1999). \\
 5. \>     \> M Razavy, Am J Phys {\bf 48}, 285(1980). \\
 6. \> (a) \> F M Arscott,Periodic Differential Equiations(Pergamon,1990);  \\
    \> (b) \> W Magnus and S Winkler, Hill's Equation(Wiley,1996). \\
 7. \> (a) \> A V Turbiner,J Phys {\bf A22} ,L1(1989) ;  \\ 
    \> (b) \> F Finkel,A Gonzalez-Lopez and M A Rodriguez, ibid {\bf A32},6821 (1999). \\
 8. \>     \> F Finkel,A Gonzalez-Lopez and M A Rodriguez, J Phys {\bf A33}, 1519(2000). \\
 9. \>     \> M A Shifman, Int J Mod Phys {\bf A4}, 2897(1989). \\
10. \>     \> C M Bender and G V Dunne, J Math Phys {\bf 37}, 6(1996). \\
11. \>     \> A Ganguly, in preparation. 
 \end{tabbing}
\end{document}